\documentclass[aps,pre,onecolumn,notitlepage,superscriptaddress]{revtex4-1}
\usepackage[utf8]{inputenc}
\usepackage[T1]{fontenc}

\usepackage{amsmath}
\usepackage{amssymb}
\usepackage{graphicx}
\usepackage{hyperref}
\usepackage[arrow, matrix, curve]{xy}
\usepackage{bm}
\usepackage{yhmath}
\usepackage{color}

\renewcommand{\vec}[1]{\mathbf{#1}}

\begin{document}

\title{Adaptive active Brownian particles searching for targets of unknown positions}

\author{Harpreet Kaur}
\affiliation{Institut f\"ur Theoretische Physik, Universit\"at Innsbruck, Technikerstra{\ss}e 21A, A-6020, Innsbruck, Austria}
\author{Thomas Franosch}
\affiliation{Institut f\"ur Theoretische Physik, Universit\"at Innsbruck, Technikerstra{\ss}e 21A, A-6020, Innsbruck, Austria}
\author{Michele Caraglio}
\email{Michele.Caraglio@uibk.ac.at}
\affiliation{Institut f\"ur Theoretische Physik, Universit\"at Innsbruck, Technikerstra{\ss}e 21A, A-6020, Innsbruck, Austria}


\begin{abstract}
Developing behavioral policies designed to efficiently solve target-search problems is a crucial issue both in nature and in the nanotechnology of the 21st century.
Here, we characterize the target-search strategies of simple microswimmers in a homogeneous environment containing sparse targets of unknown positions. 
The microswimmers are capable of controlling their dynamics by switching between   Brownian motion and an active Brownian particle and by selecting the time duration of each of the two phases.
The specific conduct of a single microswimmer depends on an internal decision-making process determined by a simple neural network associated with the agent itself.
Starting from a population of individuals with random behavior, we exploit the genetic algorithm NeuroEvolution of Augmenting Topologies to show how an evolutionary pressure based on the target-search performances of single individuals helps to find the optimal duration of the two different phases.
Our findings reveal that the optimal policy strongly depends on the magnitude of the particle's self-propulsion during the active phase and that a broad spectrum of network topology solutions exists, differing in the number of connections and hidden nodes.
\end{abstract}


\maketitle

\section*{Introduction}

Active matter and directed motion have come under the spotlight of several research communities such as biology, biomedicine, robotics, and statistical physics~\cite{Cates2012,Marchetti2013,Bechinger2016,Fodor2016,Fodor2018,Caraglio2022}.
In nature, many micro-organisms are able to convert chemical energy into self-propulsion with the goal of exploring their environment, foraging nutrients, or running away from toxic substances~\cite{Marchetti2013,Elgeti2015,Bechinger2016}.
Paradigmatic examples include the swimming behavior of bacteria such as \textit{Escherichia coli}~\cite{Berg2004}, phagocytes of the immune system performing chemotactic motion during injury or infection~\cite{Devreotes1988,Deoliveira2016}, and sperm cells navigating against chemical gradients to find the egg~\cite{Eisenbach2006}.
At larger length-scales, animals constantly have to face challenges such as finding food, a mating partner, or shelter~\cite{Viswanathan2011,Benichou2011}, and generally solve these issues by adopting strategies involving smart motion.
Artificial and biohybrid microswimmers~\cite{Elgeti2015,Smanski2016,You2018,Klumpp2019} capable of intelligent self-propulsion have potential for revolutionary applications ranging from active drug delivery~\cite{Naahidi2013,Patra2013,Cheang2014,Liu2016} to assisted fertilization~\cite{Medina-Sanchez2016} and environmental remediation~\cite{Gao2014}.

Notwithstanding the tremendous progress that has been achieved in this research field in the past decade, central problems as those regarding optimal navigation and target-search strategies still have to be thoroughly addressed already at the level of a single agent in a homogeneous environment.
Nature has found solutions to these problems in hundreds of millions of years of evolution:
Many organisms display robust locomotion performances by adapting their locomotory gaits to the surroundings~\cite{Maladen2009,Fang-Yen2010} and several microswimmers sense environmental stimuli and exhibit various tactics to achieve effective navigation in biological fluids~\cite{Elgeti2015,Berg2004,Devreotes1988,Deoliveira2016,Eisenbach2006}.
In a certain sense, even relatively simple microswimmers are smart with respect to the actions they have to perform to reach their biological goals.

To understand how evolution shaped navigation and search strategies, one can use reinforcement learning (RL)~\cite{Sutton2018} and genetic algorithms~\cite{Davis1991,Mitchell1998} to identify optimal and alternative strategies.
Recently it has been demonstrated how agents trained with RL (eventually combined with genetic algorithms) are able to find advantageous swimming strategies in several situations such as in viscous solutions~\cite{Muinos-Landin2018,Tsang2020,Hartl2021}, simple energy landscapes~\cite{Schneider2019}, steady flows~\cite{Colabrese2017,Gustavsson2017,Colabrese2018}, turbulent fluids~\cite{Reddy2016,Reddy2018,Biferale2019,Alageshan2020}, and complex motility landscapes~\cite{Monderkamp2022}.
Notwithstanding their merits, in all these studies, either the goal of the particle is different from reaching a specific target or, if a target region has to be met, its position is fixed and then implicitly learned during the learning process.
On the other hand, a crucial question arising in the context of target search is which strategies are optimal to find sparse targets of unknown positions.

The first theoretical studies on search strategies date back to World War II, when the U.S. Navy tried to rationalize search procedures to efficiently hunt submarines of the enemy~\cite{Champagne2003}.
When looking for sparse small targets, depending on the searcher's abilities and on the space to be explored, different target-search strategies can be put in place.
In the microscopic world, often the agents have only limited or no spatial memory, and search trajectories can be qualified as stochastic, meaning that some characteristics of the stochastic motion typical at this length scale are tuned to optimize the search time~\cite{Benichou2011}.
Among random strategies that can be used in a homogeneous environment, L\'{e}vy walks~\cite{Viswanathan1999,Viswanathan2008,Viswanathan2011} and intermittent-search strategies~\cite{Benichou2011,Benichou2005,Benichou2006} have been extensively studied in various contexts.
In particular, intermittent-search strategies rely on the experimental observation that fast movement degrades perception.
Thus, these strategies combine phases of diffusive motion allowing target detection and phases of ballistic motion with random orientation that allow moving quickly to a different space region but do not allow detecting the target.
It has been shown that the mean search time of intermittent random walks can be minimized under broad conditions~\cite{Loverdo2009,Benhamou1992,Moreau2009}.
Recently, Mu\~{n}oz-Gil and coworkers~\cite{munozgil2023} have proposed the use of RL techniques to study non-intermittent-search strategies and showed how these can outperform Lévy walks.

Here, for the first time, within the framework of intermittent-search strategies, we address the problem of finding a target of unknown position using a genetic-algorithm approach.
This is elaborated for the simple case of a homogeneous environment and considering agents equipped with a simple artificial neural network (ANN) which selects, based on the agent state, an action among a set of possible actions.
Specifically, the agent can switch its state between a passive and an active Brownian particle and, with intermittent-search strategies in mind, has at its disposal only a limited set of actions allowing it to switch between passive and active motion and to choose the duration of the new phase.
In our framework, the agent behavior is then deterministically selected on the basis of its state.
Our goal is to characterize and understand the behavioral policies which are optimal in solving the target search problem and show that these strategies can be obtained by means of genetic algorithms.
A genetic algorithm is here preferred over typical RL methods because having a target with a completely unknown position results in a very sparse reward signal to be used in the latter methods.
Given the pivotal role played by the reward function in RL, this is a non-trivial problem to face when willing to adopt such approach~\cite{Sutton2018}.
On the other hand, genetic algorithms are known to be less sensitive to issues related to the sparseness of the rewards because they evaluate the full behavior of an agent rather than trying to find the value of the state-action pairs.

\section*{Model}

Our environment consists of a two-dimensional square box of size $L \times L$ with periodic boundary conditions and of a circular target of radius $R$ placed randomly inside this box.
Note that, due to the periodic boundary conditions, this environment is equivalent to an infinite domain with a lattice of targets.
We consider an agent able to switch its state $s$ between a passive Brownian particle (BP) and an active Brownian particle (ABP).
Similarly to intermittent-search strategies, during the BP phase the agent is allowed to find the target while, in the ABP phase, it can more quickly relocate to a different region of the box but cannot sense the target.
Every time the agent finds a target (i.e. the distance between its position and the center of the target is smaller than the target radius $R$), this is destroyed and a new target appears at a new random location inside the box.
The equations of motion of the ABP model in a homogeneous environment are a set of Langevin equations that, once discretized according to It\^{o} rule, read
\begin{eqnarray}\label{eom1}
\vec{r}_{t\!+\!\Delta t} &=& \vec{r}_{t} + v \, \vec{u}_{t} \, \Delta t  + \sqrt{2D\Delta t} \, \boldsymbol{\xi}_t \; , \\ 
\label{eom2}
\vartheta_{t\!+\!\Delta t} &=& \vartheta_{t} + \sqrt{2D_{\vartheta}\Delta t} \, \eta_t \; ,
\end{eqnarray}
where $\Delta t$ is the integration step, $\vec{r}_t = (x_t,y_t)$ is the position at time $t$, and $\vec{u}_{t} = \big(\cos\vartheta_{t},\sin\vartheta_{t}\big)$ denotes the instantaneous orientation of the driving velocity with constant modulus $v$.
$D$ and $D_{\vartheta}$ are the translational and rotational diffusion coefficients, respectively.
Finally, the components of the vector noise $\boldsymbol{\xi}_{t}=(\xi_{x,t},\xi_{y,t})$ and of the scalar noise $\eta_t$ are independent random variables, distributed according to a Gaussian with zero average and unit variance.
The equations of motion of the standard Brownian particle model are readily recovered by setting $v=0$, thereby decoupling the spatial evolution from the orientational diffusion of the self-propulsion.
In the following, we fix the length unit as the size of the square box $L$ and time unit as the typical time $\tau := L^2/4D$ required by a passive particle to cover this distance. 
The magnitude of the activity and the persistence of motion in the ABP phase are respectively measured by the dimensionless P\'{e}clet number, $\text{Pe} := v\tau/L$ and the dimensionless persistence $\ell^* := v/D_{\vartheta}L$.

\begin{figure}[ht]
\centering
\includegraphics[width=\linewidth]{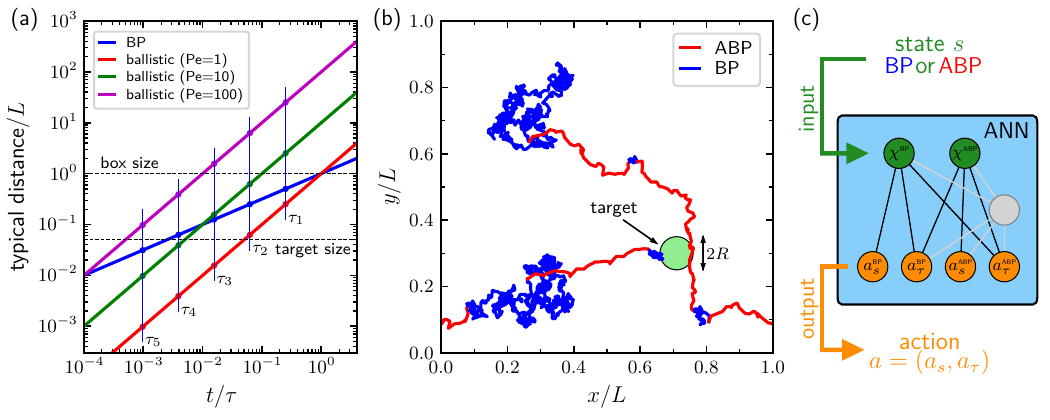}
\caption{
\textbf{(a)} Typical distance travelled during action time $\tau_i = \tau/4^i$, ($i=1,\ldots,5$) by a BP and by a particle moving ballistically for three different P\'eclet numbers. 
\textbf{(b)} Sketch of a typical trajectory exploring space by alternating between the two different phases (BP and ABP) and ending with encountering the target.
\textbf{(c)} The agent is modeled as a neural network taking as input the state of the agent itself and returning as output the action to be performed. Further details in Methods section. Each action first changes (or maintains inalterate) the state of the particle and then integrates the equations of motion for a time $\tau_i$.
}
\label{fig:model}
\end{figure}

With the intent of keeping the setup as simple as possible, we equip our particle with only a limited set actions for each state.
Each action $a= (a_s,a_{\tau})$  is a tuple consisting of two parameter. 
The first parameter, $a_s \in \{ 0, 1\}$, is a binary variable determining the next state of the particle: 
For $a_s=0$ the next phase is identical to the previous one (BP$\rightarrow$BP or ABP$\rightarrow$ABP), while for $a_s=1$ the phase changes (BP$\rightarrow$ABP or ABP$\rightarrow$BP).
The second parameter, $a_{\tau}$, specifies the time duration of the next phase and here we restrict the agent to select only among $N_{\tau} = 5$ different time durations $\tau_i$ ($i=1,\ldots,5$), thus implying a total of $10$ possible actions to select given the state.
When an active phase begins after a passive one, the direction $\vartheta$ of the self-propulsion velocity is drawn from a uniform distribution in $[0, 2\pi)$, otherwise it is updated according to Eq.~\eqref{eom2}.
In the following we set the persistence $\ell^*=1$ and select the time duration of the actions such that $\tau_i = \tau/4^i$, see Figure~\ref{fig:model}a.
Thus, in the longest action time, $\tau_1 = \tau/4$, a Brownian particle covers a typical distance of $L/2$ which, because of periodic boundary conditions, is comparable to the possible maximal distance between the particle and the target, $L/\sqrt{2}$.
The results reported in this manuscript are obtained by following this choice. However, to check that our results are not particular to this choice, the Supplemental Material reports also results obtained by varying the number of allowed time durations $N_{\tau}$.

To learn optimal policies solving the target-search problem, we exploit the evolutionary algorithm NeuroEvolution of Augmenting Topologies (NEAT)~\cite{Stanley2002,Lang2021}.
Therefor we equip each agent with an ANN taking as input the current state (BP or ABP) and outputs an action chosen among the set of actions described above, see Figure~\eqref{fig:model}c.
Any ANN is characterized by a certain topology, internal parameters, and activation and response functions, see section Methods for further details.
Depending on its inner structure, a given ANN always returns the same action given the input state.
Starting from a population of randomly generated individuals (ANNs), the NEAT algorithm then iteratively creates new generations relying on biologically inspired operators such as mutation, crossover, and selection based on the fitness of each individual in the population.
In very simple words, only the inner traits of the fittest individuals are transmitted from a generation to the next generation.
Finally, the fitness of an individual is defined as the number of targets that it manages to detect in a time equal to $5 \cdot 10^2 \tau$, see section Methods for further details.

\section*{Results}

We start by investigating how the adaptive particles evolve in the case where the size of the target is in between the typical distance explored in the passive phase in time $\tau_5$ and the typical distance covered in the active phase in the same time, $ \sqrt{\tau_5/\tau} < R/L < \text{Pe} \, \tau_5/\tau$, see Figure~\ref{fig:model}a.
Within this choice, independently of the phase duration time $\tau_i$, during the active phase the particle is typically relocating to a distance larger than the target size.
On the other hand, when the action duration $\tau_5$ is selected, a particle in the passive phase is typically exploring a region smaller than the target size.
This situation, representing well the idea of an intermittent search, is here implemented by setting the radius of the target and the P\'{e}clet number to $R=0.05L$ and $\text{Pe}=100$ respectively.

The initial population contains $N=10^3$ randomly generated individuals which, depending on the particular action returned when being in a given state, can be categorized into three different species: 
Individuals that always behave as a passive particle (BP-like individuals), those that once they visit their active state are unable to go back to the passive one (ABP-like individuals), and finally individuals that switch periodically between the two phases with fixed switching times (switching individuals).
The latter species can be further split into $N_{\tau}^2$ sub-species depending on the specific duration of the BP and the ABP phases.

\begin{figure}[ht]
\centering
\includegraphics[width=\linewidth]{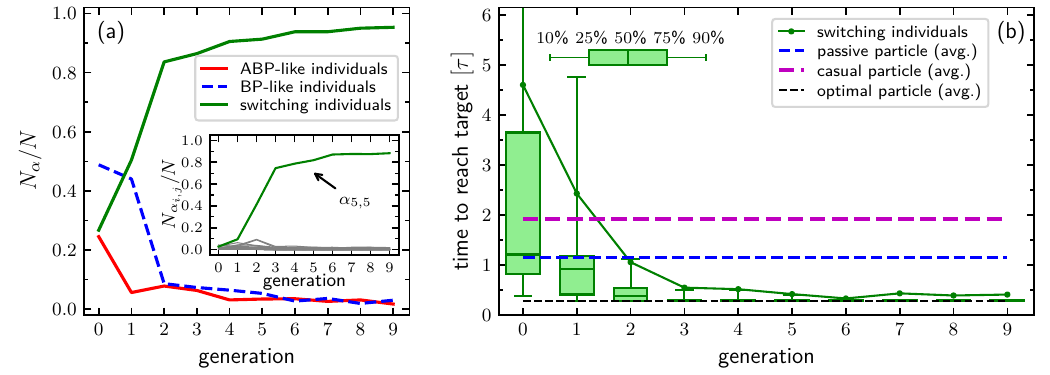}
\caption{For $R=0.05L$, $\text{Pe}=100$, and $\ell^*=1$:
\textbf{(a)} Fraction of individuals in each species $\alpha$ as a function of the generation. The inset reports the fraction of individuals in each sub-species $\alpha_{i,j}$ with $\alpha_{i,j}$ referring to switching individuals having a BP phase duration $\tau_i$ and an ABP phase duration $\tau_j$.
\textbf{(b)} Averaged time required to reach the target as a function of the generation.
Continuous line with circular points represents the average over the switching individuals, while the box-and-whiskers symbols report respectively the $10$th, $25$th, $50$th, $75$th, and $90$th percentiles among the switching individuals only.
Starting from the $3$rd generation, the $10$th, $25$th, $50$th, and $75$th percentiles are squeezed over the optimal particle benchmark.
}
\label{fig:learningPe100}
\end{figure}

Since the NEAT algorithm dynamically produces new generations by selecting for reproduction those individuals having the largest fitness, for this large value of the P\'{e}clet number, we expect that the switching individuals become the dominating species.
Indeed, the fraction of switching individuals raises from the initial value of about $0.25$ to about $0.9$ already at the second generation and reaches a plateau at about $0.95$ at the fifth generation, see Figure~\ref{fig:learningPe100}a.
Further inspection reveals that the large majority of the switching individuals is represented by those individuals that select $\tau_5$ as the time duration of both the passive and the active state, corresponding to short passive detection phases alternated to short active relocation ones, see inset of Figure~\ref{fig:learningPe100}a.

We evaluate the learning performances also by inspecting the evolution of the time required by the agent to reach the target in subsequent generations and comparing it to the same quantity computed for three different benchmarking models:
$i)$ A fully passive particle; 
$ii)$ A particle which selects randomly among the possible actions (we will refer to this benchmark as the ``\textit{casual particle}'');
$iii)$ An ``\textit{optimal particle}'' following the optimal strategy. 
This optimal strategy is obtained by individually checking the $N_{\tau} \times N_{\tau}$ possible combinations that make the particle switch to the active state when it is in the passive phase and to the passive state when it is in the active one.

At the beginning of the learning process, for the switching individuals, the time required to find the target is comparable to that of the casual particle.
More precisely, the average searching time of the switching individuals is twice that of the casual particle, while the median is about a factor of $0.6$ lower than the latter quantity and closer to the searching time of a completely passive particle.
This observation is due to the fact that the phase times $\tau_i$ ($i=1,\ldots,5$) are logarithmically spaced and that, in the initial population of switching individuals, some spend most of their time in the active phase where it is not possible to find the target, and others spend most of their time in the passive phase, thus effectively performing as a Brownian particle.
If the evolutionary process is successful, after repeated generations this quantity is expected to decrease and achieve a value slightly bigger than that of the optimal particle, with this small difference due to the fact that the NEAT algorithm maintains a certain level of exploration by creating new individuals through mutations and preserving a certain number of unfit individuals for different evolution scenarios.
This is in fact what is observed for $R=0.05L$ and $\text{Pe}=100$, see Figure~\ref{fig:learningPe100}b.
Furthermore, since more and more switching individuals enter the most adapted subspecies, also the spread of the search time decreases during the learning process, with at least $75$\% of the switching individuals behaving as good as the optimal particle after the third generation, see Figure~\ref{fig:learningPe100}b.
However, the average stays above the $75$th percentile over the whole evolutive process because is dominated by the performances of the worst switching individuals. 

\begin{figure}[ht]
\centering
\includegraphics[width=\linewidth]{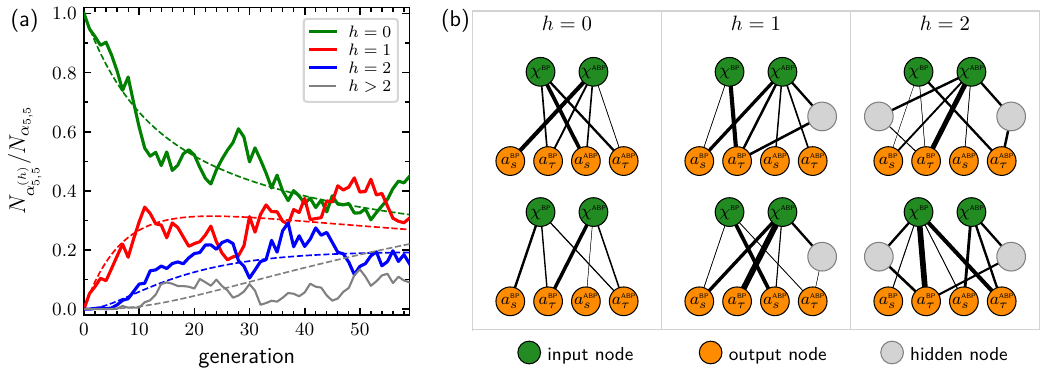}
\caption{For $R=0.05L$, $\text{Pe}=100$, and $\ell^*=1$:
\textbf{(a)} Fraction of individuals with different amounts of hidden nodes, $h$, in subspecies $\alpha_{5,5}$ (individuals having both the BP and the ABP phase lasting $\tau_5$) as a function of the generation.
Dashed lines are the theoretical expectations obtained by solving the master equations for each group (probability of adding or deleting a node is $0.05$).
\textbf{(b)} Typical ANN topologies for individuals belonging to the subspecies  $\alpha_{5,5}$ and with a number of hidden nodes equal to $h=0$, $h=1$, and $h=2$, respectively. Individuals are extracted from the $9$th generation. The width of the edges is proportional to the absolute value of the weight of the connections between different nodes.
}
\label{fig:topologyPe100}
\end{figure}

Additional insight into how the genetic algorithm encodes learning a successful policy can be obtained by focusing on the topology of the ANNs corresponding to the fittest individuals, i.e. for $\text{Pe}=100$, those selecting $\tau_5$ as the duration of both the BP and ABP phases (inset of Figure~\ref{fig:learningPe100}a).
In the initial population, all the individuals have two input nodes and four output nodes with no hidden nodes in between, see Methods section for details.
However, with the progress of the evolution process, new fit individuals with some hidden nodes emerge, see Figure~\ref{fig:topologyPe100}a.
The fractions of fit individuals with $h$ hidden nodes are consistent with the expected values as obtained by solving the master equation
\begin{equation}
N_h (g) = N_h(g-1) - p_{\rm add} \, N_h (g-1)  - p_{\rm del} \, N_h (g-1)  + p_{\rm add} \, N_{h-1} (g-1)  + p_{\rm del} \, N_{h+1}(g-1)  \; ,
\end{equation}
where $N_h(g)$ indicates the number of individuals with $h$ hidden nodes at the $g$-th generation and $p_{\rm add}=p_{\rm del}=0.05$ are respectively the probability of adding and deleting an hidden node.
This means that the topology of the initial ANN with only two input nodes and four output nodes is already complex enough to provide successful solutions to the target-search problem.
In contrast, a similar approach adopted starting from a different initial network topology shows that, when having fewer output nodes, fittest individuals are slightly likely to have a certain number of hidden nodes that contribute in selecting the optimal actions, see Supplemental Material.
This observation agrees with the intuitive expectation that a minimal ANN's complexity is required to find successful strategies.

By looking more in detail at the internal structure of the ANNs associated with the fittest individuals, it appears that there is not a preferred topology once the number of hidden nodes $h$ is fixed.
In fact, typical topologies of these networks show a very variagated range of active connections and of their weights among the different nodes, see figures~\ref{fig:topologyPe100}b and~S2 in the Supplemental material.
In these plots the edges connecting different vertices have a width proportional to the absolute value of the weight of the connection, see Methods section for details.
However, a complete understanding of the functioning of the particular ANN should include also the properties of the single nodes, including their bias and activation and aggregation functions, see Methods section for details.
Note also that, typically, emergent hidden nodes are connected to only one input and one output.

\begin{figure}[ht]
\centering
\includegraphics[width=\linewidth]{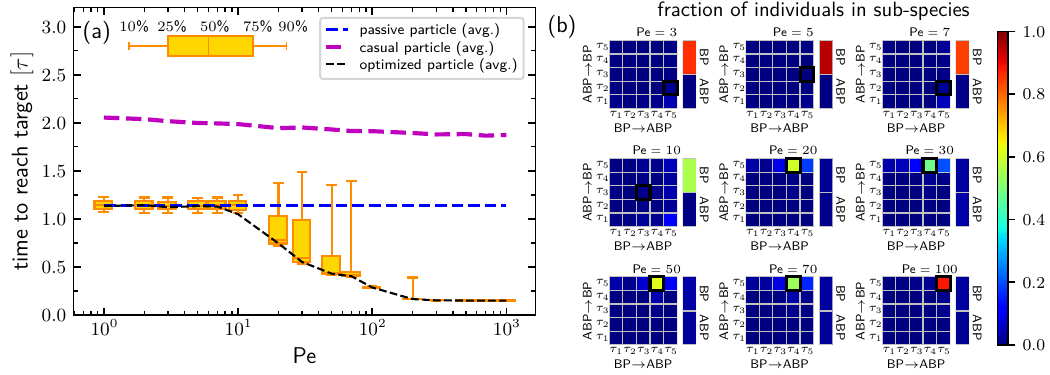}
\caption{For $R=0.05L$ and $\ell^*=1$:
\textbf{(a)} Time required to reach the target as a function of the P\'eclet number at $9$-th generation.
The box-and-whiskers symbols report respectively the $10$th, $25$th, $50$th, $75$th, and $90$th percentiles among the whole population.
The dashed lines represent the values for the three benchmarks.
\textbf{(b)} Fraction of individuals in each sub-species for different P\'eclet numbers. The box highlighed in black represents the action selected by the optimal particle.}
\label{fig:learning_various_Pe}
\end{figure}

An important question is how the target-search strategy depends on the activity of the particle.
To address this issue, we carry out a similar analysis for a different P\'eclet numbers.
When $\text{Pe}>100$ we do not expect major differences to the previously considered case ($\text{Pe}=100$).
In fact, for high P\'eclet numbers, the self-propulsion velocity is enough to allow relocating at a distance larger than the target size even in the shortest time $\tau_5$.
Then, intuitively, performing a motion alternating between short BP phases and short ABP phases is the most promising strategy.
In contrast, for very low P\'eclet numbers, if the action time is smaller then the time unit $\tau$, the typical distance travelled by simple diffusion is always greater than the distance covered by self-propulsion, see Figure~\ref{fig:model}.
Then, selecting active phases becomes superfluous and the optimal strategy is simply maintaining the particle in a BP-like phase.
These expectations are confirmed by checking, after about $10$ generations, the fraction of individuals in different subspecies and by comparing the overall performance of the population to those of the passive and of the optimal particle for different P\'eclet numbers.
This is done in Figure~\ref{fig:learning_various_Pe}a, reporting the average search times, and in Figure~\ref{fig:learning_various_Pe}b, showing the  relative distribution of individuals across subspecies. In the latter panel, the phase durations associated with the optimal particle are also highlighted with a black frame in the table.

More interesting is the behavior for intermediate activity ($10 \lesssim \text{Pe} \lesssim 70$).
In this range, the optimal particle outperforms the simple passive particle and the optimal actions correspond to a behavioral policy that again displays the shortest passive phase (i.e. BP-like phases with duration $\tau_5$) but, to allow for significant relocation, selects a longer duration of the ABP phase, namely the duration $\tau_2$ for the considered cases, see Figure~\ref{fig:learning_various_Pe}b.
Consistently, for $\text{Pe}=20,30,50,$ and $70$, the population produced by NEAT at the $9$-th generation shows a majority of individuals belonging to the sub-species selecting the same actions as those selected by the optimal particle.
However, for these values of activity, the genetic algorithm still considers as possible candidates for the optimal solution also individuals of other sub-species, especially those with a short passive phase.
This results in the higher spread of the search times reported in Figure~\ref{fig:learning_various_Pe}a.
Finally, at $\text{Pe}=10$ the optimal solution corresponds to an optimal particle selecting $\tau_3$ as the duration of both the passive and the active phase but the NEAT algorithm in this case develops a population with a majority of BP-like individuals.
The reason for this small inconsistency is likely due to the fact that, in this case, the average search time of the passive particle is very close to that of the optimal particle and, by chance, the first evolutionary lineage is preferred.

A final remark is in order: All results reported so far are obtained by strongly reducing the amount of possible ABP and BP phase durations.
In particular, we allowed only $N_{\tau}=5$ different durations that span a large time range.
By doing so, the computational time required to run the NEAT algorithm is comparable with directly evaluating the performances of the completely Brownian particle and of the $N_{\tau} \times N_{\tau}$ possible switching individuals.
However, any integer multiple of the integration time step $\Delta t$ could serve as a possible phase duration and one may consequently increase $N_{\tau}$, thus letting the ANNs select phase durations among a more fine-grained palette.
Indeed, increasing $N_{\tau}$ is desiderable to explore a larger set of switching individuals, thus allowing to find even better phases durations that increase the target-search performances.
However, by doing so, a brute force check of all possible switching individuals becomes quickly inefficient, with computational costs increasing as $N_{\tau}^2$, in favor of our choice of using the NEAT algorithm, which, in contrast, has unaltered computational costs.
The Supplemental Material reports the equivalent of figures~\ref{fig:learningPe100} and~\ref{fig:learning_various_Pe} obtained by varying $N_\tau$, namely for $N_\tau=10$, $20$, and $50$.
The learning performances obtained by the NEAT algorithm in these cases are comparable to the case study reported in the main text and, for large P\`eclet numbers, the average time to reach the target slightly outperforms that of the fittest individual when this is selected among the set of $25$ switching individuals used in this section, see Figure S6 and S8 in the Supplemental Material.
However, the fluctuations of the average time to reach the target increase with $N_{\tau}$.
Furthermore, with increasing number of allowed phase durations, more actions configure themselves as a valid candidate to become the optimal action selected by the genetic algorithm.

\section*{Conclusions}

Our findings demonstrate that genetic algorithms are a powerful tool to address the problem of finding targets of unknown positions for particles able to switch their behavior between a simple passive Brownian particle and an active Brownian particle.
In particular, we equipped the particle with a neural network receiving the current state of the particle itself as the only input and returning as an output a decision regarding if and when the agent should switch its phase.
We then showed that the algorithm NeuroEvolution of Augmenting Topologies is able to evolve an initial population of neural networks taking random decisions towards a population in which the majority of individuals are optimized to solve the target-search problem.

In principle, similar results on target-search performances of intermittent passive-active Brownian particles could be obtained by resorting on RL algorithms~\cite{Sutton2018}.
However, in the RL framework, setting a target having a different unknown position at the beginning of each target-search episode would result in a very sparse reward function, which is generally a non-trivial problem in RL~\cite{Sutton2018}.
More specifically, since in our case the state of the agent is a simple binary variable and the rewards are extremely sparse, the reward signal has only a very low correlation with the particular state-action pair encountered when the target is found, making typical action-value methods such as Q-learning or SARSA\cite{Sutton2018} fail in learning successful strategies.
If willing to follow any RL method, algorithms taking into account long sequences of visited state-action pairs should then be preferred because these methods, similar to the genetic algorithm, seek to maximize the performances by directly evaluating the outcome of a given policy. Possible algorithms of this kind include policy-gradient and actor-critic methods\cite{Sutton2018} and the projective simulations algorithm~\cite{Briegel2012}.
These considerations do not exclude that successful results could be obtained by using more elaborated versions of the above mentioned action-value methods and/or by differently defining the states and the actions, as proposed in Ref.~\cite{munozgil2023} for non-intermittent searchers.
On the other hand, genetic algorithms bypass the problem of reward sparseness by giving the agents a fitness function dependent on the overall performance in assessing some task, and thus they are particularly suited to our case.

In the current setup, the output of any given individual ANN is fixed once the input is given, meaning that, given its current state, a particle always chooses deterministically the same action.
This is different from the typical intermittent-search strategies discussed in Ref.~\cite{Benichou2011}, where an agent draws the phase durations from a certain distribution.
However, similarly to the most general case~\cite{Benichou2011,Loverdo2009}, also our results show that there is an optimal duration of the active relocation phase which depends mainly on the amount of the activity and that, for very low activity, having an active phase is not any more functional to improve the target-finding efficiency.
A natural step forward to go even further in the direction of a standard intermittent search model would be to adapt our algorithm in such a way that, instead of learning the optimal phase durations, it optimizes the parameters of a certain time distribution.
Another possible extension of our work would be to link the output of the ANNs to a transition probability rather than to a deterministic action, with different individuals thus corresponding to different transition matrices.

Our paper provides a first attempt to use machine-learning methods to investigate the problem of finding targets of unknown positions in a simple homogeneous environment and paves the way to further research on this important problem.
Having a minimal model with only two distinct phases is a choice that serves as our proof of concept that genetic algorithms are powerful tools to investigate target-search strategies in stochastic systems.
However, in nature, searchers may have multiple dynamic modes.
For example, dendritic cells searching for infections combine three distinct migration modes in their motion~\cite{Song2023} and some DNA-binding proteins also have more than two dynamic states for the search~\cite{Halford2004}.
This provides a solid biological motivation for a first generalization of our work to a case in which three or more distinct different phases are considered as possible states of the agent.
Other possible topics worth future investigation include multiple and/or motile targets problems~\cite{Benichou2011}, target search with resetting events~\cite{Evans2011,Kusmierz2014,Kumar2020}, and extensions to more realistic scenarios involving the presence of boundaries, obstacles, and energy barriers~\cite{Volpe2017,Zanovello2021,Zanovello2021b,Zanovello2023}.
To follow these goals, both the state and the actions of the agent may be made arbitrarily complex: For example, the agent could gather sensorimotor cues from the environment and, based on them, modify its behavior by control over some motility parameters.
Alternatively, similarly to what has been done recently in a different context~\cite{munozgil2023}, the agent can be equipped with the ability to sense the duration of its current phase.
Finally, agents having a limited memory of the visited locations can also be investigated.

\section*{Methods}

To investigate how an evolutionary pressure allows the adaptive particles to develop successful target-search strategies, we resort to the genetic algorithm NeuroEvolution of Augmenting Topologies (NEAT)~\cite{Stanley2002,Lang2021,NEAT}.
To do so, a simple Artificial Neural Network (ANN) is associated with our adaptive particle.
The role of this network is to take as input the state of the particle ($s=$BP or ABP) and return as output the action to be performed as described in the Model section.
Starting from a population of $10^3$ individuals, the NEAT algorithm then iteratively creates new generations relying on biologically inspired
operations such as mutation, crossover, and selection based on the fitness of each individual in the population.
This fitness is defined as the number of targets that the individual manages to detect in a time equal to $5 \cdot 10^2 \tau$.
The evolution process is based on the principle of complexification of existing networks~\cite{Stanley2002}: not only the node biases and the edge weights are adjusted to optimize the individuals' fitness, but this goal is also reached by changing the topology of the network, i.e. by adding or deleting new nodes and enabling or disabling some connections, see text below.
The number of individuals in each generation is kept fixed.

More specifically, we construct an initial population of networks having two input nodes and four output nodes, see Figure~\ref{fig:model}c.
These are the nodes through which the ANN interacts with the external environment and they cannot be created or destroyed.
The two input nodes recognize the state $s$ of the particle and pass to the output nodes the signals $\chi^{\rm BP}(s)$ and $\chi^{\rm ABP}(s)$ which is further multiplied by the weight of the connection between the specific pair of input and output nodes.
This signal is a simple characteristic function, $\chi^{S}(s)=1$ if $s=S$ and $\chi^{S}(s)=0$ otherwise.
The four output nodes aggregate by summation the signals coming from the various input nodes, add a bias, and return an output value according to a certain activation function.
Formally, the output value of node $j$ is given by $f(\sum_{k} w_{kj} x_k + b_j)$, where $x_k$ is the signal coming from node $k$, $w_{kj}$ is the weight of the connection between $k$ and $j$, $b_j$ is the bias of node $j$, and $f$ is the activation function, in our case a modified clamped function, $[1+f(x)]/2$ with $f(x)=-1$ if $x<-1$, $f(x)=x$ if $-1 \leq x \leq 1$, and $f(x)=1$ otherwise.
Each output node $j$ then returns an output real variable $x_j \in [0,1]$ and these four output values are then together determining the action taken by the agent as follows.
If the current state of the agent is $s=$ BP (ABP) the first (third) node determines the next state, being this again BP (ABP) if the node output value is smaller than $0.5$ or changing to ABP (BP) otherwise.
The two options correspond respectively to $a_s = 0,1$, see Model section.
The duration of the next phase, $a_{\tau}$, is instead determined by the second (fourth) output node that selects the phase duration $\tau_i$ with $i$ the integer part of $(1+x_j N_{\tau})$, being $x_j$ is the output signal of the node and $N_{\tau}=5$ the number of allowed phase durations, see Model section.

The individuals in the initial population are selected randomly (i.e. random biases and random connection weights) but all share the just described topology.
However, in subsequent generations individuals with new hidden nodes emerge.
These nodes receive the signals coming from the input nodes and eventually from other hidden nodes and, in the same fashion as described for the output nodes, return an output value which is collected as input signal from the output nodes and eventually from other hidden nodes.
During mutations, hidden nodes are generated with a probability $p_{\rm add}$ and deleted with probability $p_{\rm del}$.
Following standard practice~\cite{Lang2021,NEAT}, we set $p_{\rm add}=p_{\rm del}=0.05$.

The initial network topology with two input nodes and four output nodes works particularly well in our target-search problem.
However, while in the main text we report only results obtained by starting with this initial setup, we also tested other topologies as well as a different number of phase durations, see Supplemental Material.

\section*{References}

\providecommand{\newblock}{}

\section*{Acknowledgements}
H.K. acknowledges funding from the European Union's Horizon 2020 research and innovation programme under the Marie Skłodowska-Curie grant agreement No 847476;
M.C. is supported by FWF: P 35872-N;
T.F. acknowledges funding by FWF: P 35580-N.

\section*{Author contributions statement}
M.C. and T.F. conceived the research,  M.C. developed the software, and M.C. and H.K. analyzed the results.  All authors wrote and reviewed the manuscript.

\clearpage

\pagestyle{plain}

\begin{center}
\textbf{{\Large Supplemental Material for \\
``Adaptive Active Brownian particles searching for targets of unknown positions''}}
\medskip

Harpreet Kaur$^1$, Thomas Franosch$^1$, and Michele Caraglio$^1$ \\
$^1$\textit{Institut f\"ur Theoretische Physik, Universit\"at Innsbruck, Technikerstra{\ss}e 21A, A-6020, Innsbruck, Austria}
\end{center}

\bigskip

\section*{I. Different topology of the individuals in first generation}

Here, we consider a different choice for the initial topology of the artificial networks associated with the individuals of the first generation.

More specifically, we construct an initial population of networks having one input node and only two output nodes.
The input nodes recognizes the state $s$ of the particle and pass to the output nodes a signal $s=0$ if the particle is in the diffusive Brownian phase and $s=1$ if the particle is in the active Brownian phase.
The two output nodes are then together determining the action taken by the agent.
In particular the first node determines the next state, remaining unchanged if the node output value is smaller than $0.5$ and switching to the other state otherwise.
The two options correspond respectively to $a_s = 0,1$.
The duration of the next phase, $a_{\tau}$, is instead determined by the second output node selecting the phase duration $\tau_i$ with $i$ the integer part of $(1+x_j N_{\tau})$, being $x_j \in [0,1]$ the output signal of the node and $N_{\tau}=5$ the number of allowed phase durations, see Model section in the main text.

\begin{figure}[ht]
\centering
\includegraphics[width=\linewidth]{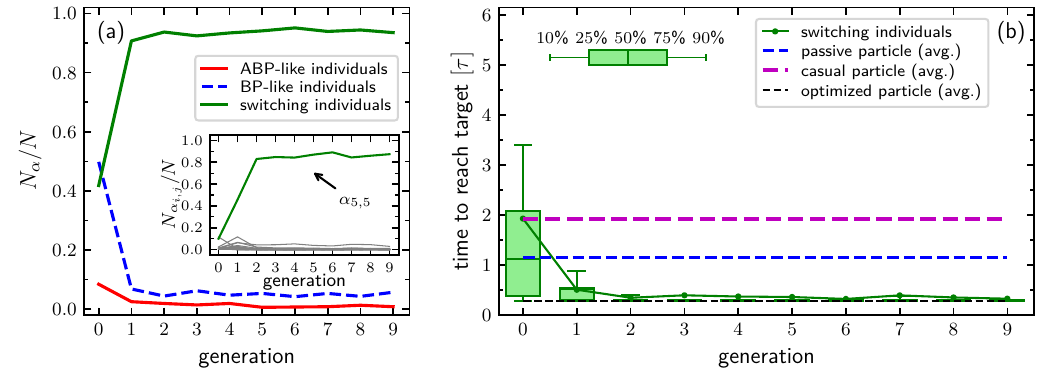}
\caption{For $R=0.05L$, $\text{Pe}=10^2$, and $\ell^*=1$:
\textbf{(a)} Fraction of individuals in each species $\alpha$ as a function of the generation. The inset reports the fraction of individuals in each sub-species $\alpha_{i,j}$ with $\alpha_{i,j}$ referring to switching individuals having a BP phase duration $\tau_i$ and an ABP phase duration $\tau_j$.
\textbf{(b)} Time required to reach the target as a function of the generation.
Continuous line with circular points represents the average over the switching individuals, while the box-and-whiskers symbols report respectively the $10$th, $25$th, $50$th, $75$th, and $90$th percentiles among the switching individuals only.
The dashed lines represent the values for the three benchmarks.
}
\label{figS1}
\end{figure}

\begin{figure}[ht]
\centering
\includegraphics[width=\linewidth]{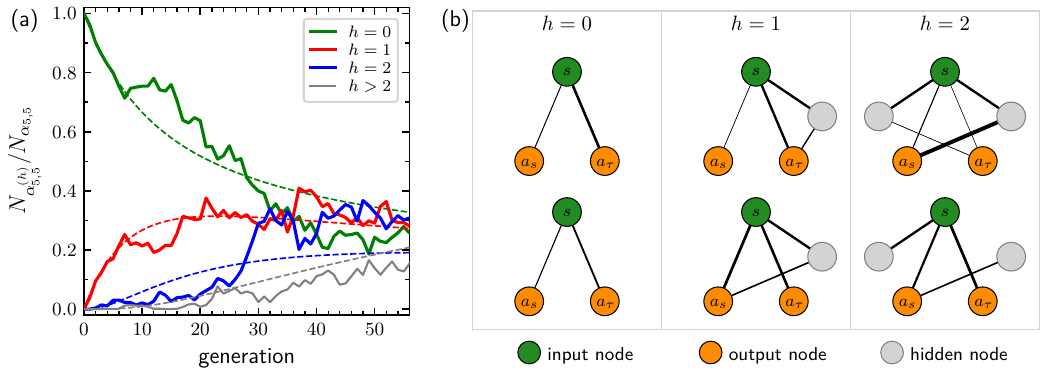}
\caption{For $R=0.05L$, $\text{Pe}=10^2$, and $\ell^*=1$:
\textbf{(a)} Fraction of individuals with different amounts of hidden nodes, $h$, in subspecies $\alpha_{5,5}$ (individuals having both the BP and the ABP phase lasting $\tau_5$) as a function of the generation.
Dashed lines are the theoretical expectations obtained by solving the master equations for each group (probability of adding or deleting a node is $0.05$).
\textbf{(b)} Typical ANN topologies for individuals belonging to the subspecies  $\alpha_{5,5}$ and with a number of hidden nodes equal to $h=0$, $h=1$, and $h=2$, respectively. Individuals are extracted from the $9$th generation. The width of the edges is proportional to the absolute value of the weight of the connections between different nodes.
}
\label{figS2}
\end{figure}

Figures~\ref{figS1} and~\ref{figS2} are respectively the analogs of figures~2 and~3 of the main text as obtained by using the previous choice for the initial network topology.

\clearpage

\section*{II. Different number of phase durations}

\subsection*{A. $N_\tau = 10$}

In this section we fix $N_\tau = 10$ and define the $\tau_i = \tau/2^i$ with $i=1,\ldots,N_\tau$, see Model section in the main text for more details.

\begin{figure}[ht]
\centering
\includegraphics[width=\linewidth]{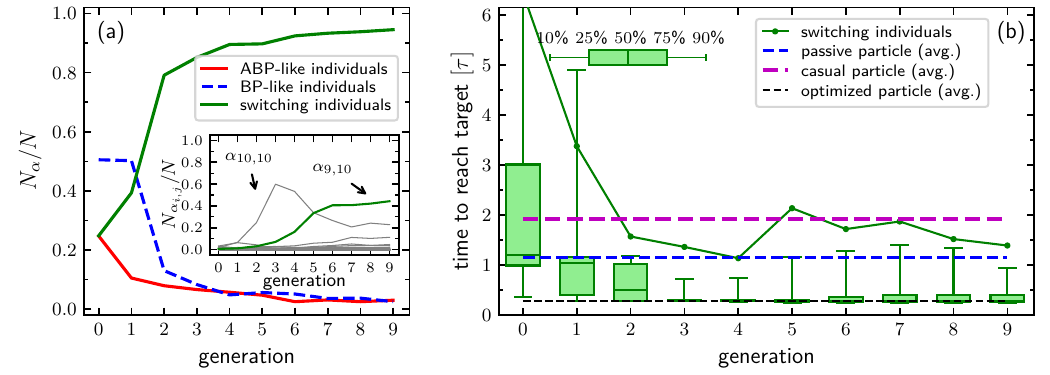}
\caption{For $R=0.05L$, $\text{Pe}=10^2$, and $\ell^*=1$:
\textbf{(a)} Fraction of individuals in each species $\alpha$ as a function of the generation. The inset reports the fraction of individuals in each sub-species $\alpha_{i,j}$ with $\alpha_{i,j}$ referring to switching individuals having a BP phase duration $\tau_i$ and an ABP phase duration $\tau_j$.
\textbf{(b)} Time required to reach the target as a function of the generation.
The dashed lines represent the values for the three benchmarks and have the same values as in Fig.2 of the main text (i.e. no brute-force individual check of the performances of the the $N_{\tau} \times N_{\tau}$ possible switching individuals has been done).
}
\label{figS3}
\end{figure}

\begin{figure}[ht]
\centering
\includegraphics[width=\linewidth]{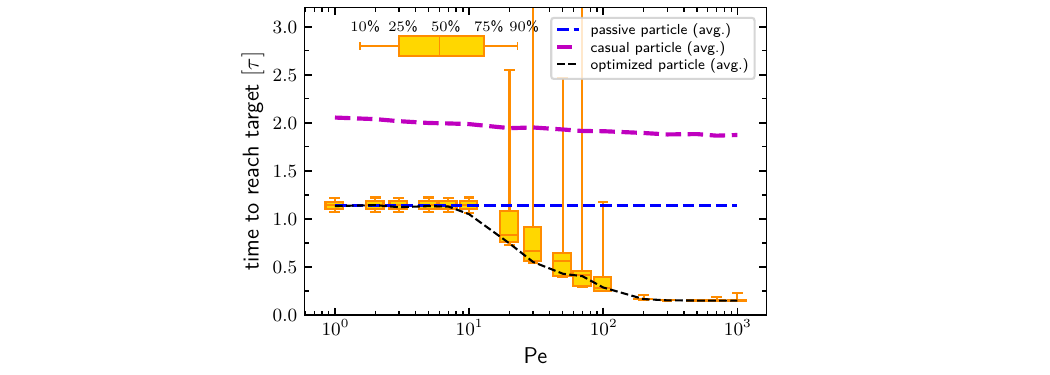}
\caption{For $R=0.05L$ and $\ell^*=1$:
Time required to reach the target as a function of the P\'eclet number at $9$-th generation.
The box-and-whiskers symbols report respectively the $10$th, $25$th, $50$th, $75$th, and $90$th percentiles among the whole population.
The dashed lines represent the values for the three benchmarks and have the same values as in Fig.4 of the main text (i.e. no brute-force individual check of the performances of the the $N_{\tau} \times N_{\tau}$ possible switching individuals has been done}
\label{figS4}
\end{figure}

\clearpage

\subsection*{B. $N_\tau = 20$}

In this section we fix $N_\tau = 20$ and define the $\tau_i = \tau/1.5^i$ with $i=1,\ldots,N_\tau$, see Model section in the main text for more details.

\begin{figure}[ht]
\centering
\includegraphics[width=\linewidth]{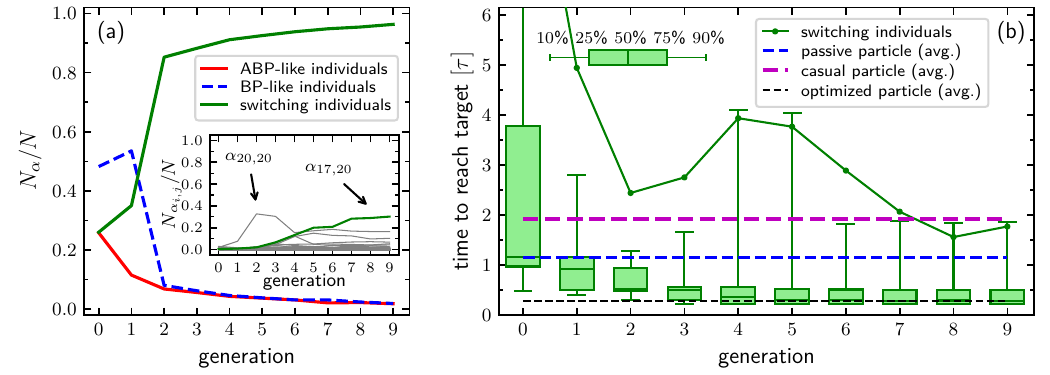}
\caption{For $R=0.05L$, $\text{Pe}=10^2$, and $\ell^*=1$:
\textbf{(a)} Fraction of individuals in each species $\alpha$ as a function of the generation. The inset reports the fraction of individuals in each sub-species $\alpha_{i,j}$ with $\alpha_{i,j}$ referring to switching individuals having a BP phase duration $\tau_i$ and an ABP phase duration $\tau_j$.
\textbf{(b)} Time required to reach the target as a function of the generation.
The dashed lines represent the values for the three benchmarks and have the same values as in Fig.2 of the main text (i.e. no brute-force individual check of the performances of the the $N_{\tau} \times N_{\tau}$ possible switching individuals has been done).
}
\label{figS5}
\end{figure}

\begin{figure}[ht]
\centering
\includegraphics[width=\linewidth]{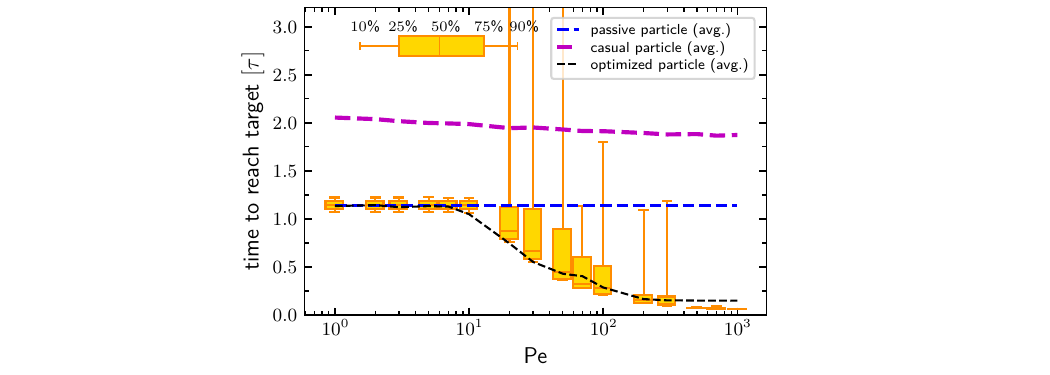}
\caption{For $R=0.05L$ and $\ell^*=1$:
Time required to reach the target as a function of the P\'eclet number at $9$-th generation.
The box-and-whiskers symbols report respectively the $10$th, $25$th, $50$th, $75$th, and $90$th percentiles among the whole population.
The dashed lines represent the values for the three benchmarks and have the same values as in Fig.4 of the main text (i.e. no brute-force individual check of the performances of the the $N_{\tau} \times N_{\tau}$ possible switching individuals has been done}
\label{figS6}
\end{figure}

\clearpage

\subsection*{C. $N_\tau = 50$}

In this section we fix $N_\tau = 50$ and define the $\tau_i = \tau/1.16^i$ with $i=1,\ldots,N_\tau$, see Model section in the main text for more details.

\begin{figure}[ht]
\centering
\includegraphics[width=\linewidth]{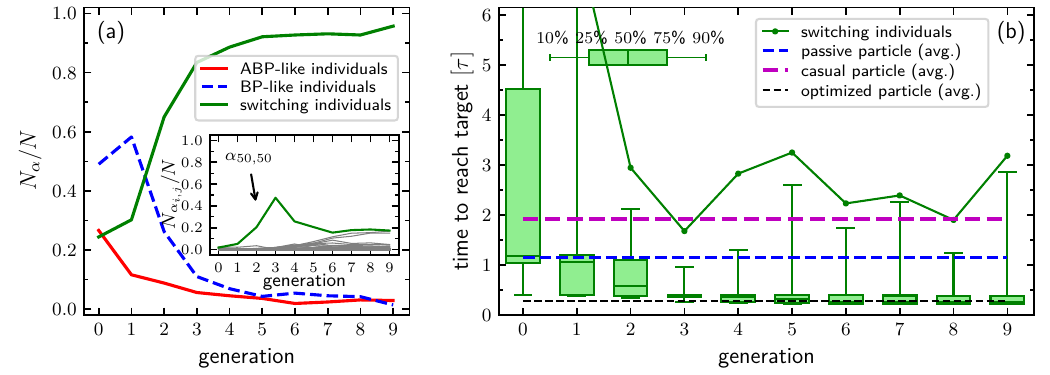}
\caption{For $R=0.05L$, $\text{Pe}=10^2$, and $\ell^*=1$:
\textbf{(a)} Fraction of individuals in each species $\alpha$ as a function of the generation. The inset reports the fraction of individuals in each sub-species $\alpha_{i,j}$ with $\alpha_{i,j}$ referring to switching individuals having a BP phase duration $\tau_i$ and an ABP phase duration $\tau_j$.
\textbf{(b)} Time required to reach the target as a function of the generation.
The dashed lines represent the values for the three benchmarks and have the same values as in Fig.2 of the main text (i.e. no brute-force individual check of the performances of the the $N_{\tau} \times N_{\tau}$ possible switching individuals has been done).
}
\label{figS7}
\end{figure}

\begin{figure}[ht]
\centering
\includegraphics[width=\linewidth]{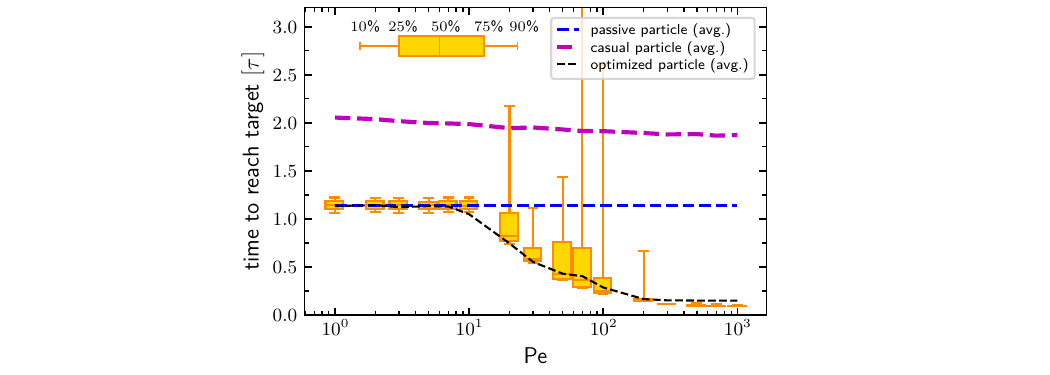}
\caption{For $R=0.05L$ and $\ell^*=1$:
Time required to reach the target as a function of the P\'eclet number at $9$-th generation.
The box-and-whiskers symbols report respectively the $10$th, $25$th, $50$th, $75$th, and $90$th percentiles among the whole population.
The dashed lines represent the values for the three benchmarks and have the same values as in Fig.4 of the main text (i.e. no brute-force individual check of the performances of the the $N_{\tau} \times N_{\tau}$ possible switching individuals has been done}
\label{figS8}
\end{figure}


\begin{thebibliography}{10}
\expandafter\ifx\csname url\endcsname\relax
  \def\url#1{{\tt #1}}\fi
\expandafter\ifx\csname urlprefix\endcsname\relax\def\urlprefix{URL }\fi
\providecommand{\eprint}[2][]{\url{#2}}

\bibitem{Cates2012}
Cates M~E 2012 {\em Reports on Progress in Physics\/} {\bf 75} 042601

\bibitem{Marchetti2013}
Marchetti M~C, Joanny J~F, Ramaswamy S, Liverpool T~B, Prost J, Rao M and Simha
  R~A 2013 {\em Reviews of Modern Physics\/} {\bf 85} 1143

\bibitem{Bechinger2016}
Bechinger C, Di~Leonardo R, L{\"o}wen H, Reichhardt C, Volpe G and Volpe G 2016
  {\em Reviews of Modern Physics\/} {\bf 88} 045006

\bibitem{Fodor2016}
Fodor {\'E}, Nardini C, Cates M~E, Tailleur J, Visco P and van Wijland F 2016
  {\em Physical Review Letters\/} {\bf 117} 038103

\bibitem{Fodor2018}
Fodor {\'E} and Marchetti M~C 2018 {\em Physica A: Statistical Mechanics and
  its Applications\/} {\bf 504} 106

\bibitem{Caraglio2022}
Caraglio M and Franosch T 2022 {\em Physical Review Letters\/} {\bf 129} 158001

\bibitem{Elgeti2015}
Elgeti J, Winkler R and G G 2015 {\em Reports on Progress in Physics\/} {\bf
  78} 056601

\bibitem{Berg2004}
Berg H 2004 {\em E. coli in Motion\/} (Springer-Verlag, Heidelberg)

\bibitem{Devreotes1988}
Devreotes P~N and Zigmond S~H 1988 {\em Annual Review of Cell Biology\/} {\bf
  4} 649

\bibitem{Deoliveira2016}
de~Oliveira S, Rosowski E~E and Huttenlocher A 2016 {\em Nature Reviews
  Immunology\/} {\bf 16} 378

\bibitem{Eisenbach2006}
Eisenbach M and Giojalas L~C 2006 {\em Nature Reviews Molecular Cell Biology\/}
  {\bf 7} 276

\bibitem{Viswanathan2011}
Viswanathan G~M, Da~Luz M~G, Raposo E and Stanley H 2011 {\em The Physics of
  foraging: {A}n introduction to random searches and biological encounters\/}
  (Cambridge University Press)

\bibitem{Benichou2011}
B\'enichou O, Loverdo C, Moreau M and Voituriez R 2011 {\em Review of Modern
  Physics\/} {\bf 83} 81

\bibitem{Smanski2016}
Smanski M~J, Zhou H, Claesen J, Shen B, Fischbach M~A and Voigt C~A 2016 {\em
  Nature Reviews Microbiology\/} {\bf 14} 135

\bibitem{You2018}
You M, Chen C, Xu L, Mou F and Guan J 2018 {\em Accounts of Chemical
  Research\/} {\bf 51} 3006

\bibitem{Klumpp2019}
Klumpp S, Lef{\'e}vre C~T, Bennet M and Faivre D 2019 {\em Physics Reports\/}
  {\bf 789} 1

\bibitem{Naahidi2013}
Naahidi S, Jafari M, Edalat F, Raymond K, Khademhosseini A and Chen P 2013 {\em
  Journal of Controlled Release\/} {\bf 166} 182

\bibitem{Patra2013}
Patra D, Sengupta S, Duan W, Zhang H, Pavlick R and Sen A 2013 {\em
  Nanoscale\/} {\bf 5} 1273

\bibitem{Cheang2014}
Cheang K~U, Kyoungwoo L, Anak~Agung J and Min~Jun K 2014 {\em Applied Physics
  Letters\/} {\bf 105} 083705

\bibitem{Liu2016}
Liu J, Wei T, Zhao J, Huang Y, Deng H, Kumar A, Wang C, Liang Z, Ma X and Liang
  X~J 2016 {\em Biomaterials\/} {\bf 91} 44

\bibitem{Medina-Sanchez2016}
Medina-S{\'a}nchez M, Schwarz L, Meyer A~K, Hebenstreit F and Schmidt O~G 2016
  {\em Nano Letters\/} {\bf 16} 555

\bibitem{Gao2014}
Gao W and Wang J 2014 {\em ACS Nano\/} {\bf 8} 3170

\bibitem{Maladen2009}
Maladen R~D, Ding Y, Li C and Goldman D~I 2009 {\em Science\/} {\bf 325} 314

\bibitem{Fang-Yen2010}
Fang-Yen C, Wyart M, Xie J, Kawai R, Kodger T, Chen S, Wen Q and Samuel A~D~T
  2010 {\em Proceedings of the National Academy of Sciences\/} {\bf 107} 20323

\bibitem{Sutton2018}
Sutton R~S and Barto A~G 2018 {\em Reinforcement {L}earning (2nd edition)\/}
  (The MIT Press)

\bibitem{Davis1991}
Davis L 1991 {\em Handbook of {G}enetic {A}lgorithms\/} (Van Nostrand Reinhold)

\bibitem{Mitchell1998}
Mitchell M 1998 {\em An introduction to {G}enetic {A}lgorithms\/} (The MIT
  Press)

\bibitem{Muinos-Landin2018}
Mui{\~n}os-Landin S, Fischer A, Holubec V and Cichos F 2021 {\em Science
  Robotics\/} {\bf 6}

\bibitem{Tsang2020}
Tsang A~C~H, Tong P~W, Nallan S and Pak O~S 2020 {\em Physical Review Fluids\/}
  {\bf 5} 074101

\bibitem{Hartl2021}
Hartl B, H{\"u}bl M, Kahl G and Z{\"o}ttl A 2021 {\em Proceedings of the
  National Academy of Sciences\/} {\bf 118} e2019683118

\bibitem{Schneider2019}
Schneider E and Stark H 2019 {\em {EPL} (Europhysics Letters)\/} {\bf 127}
  64003

\bibitem{Colabrese2017}
Colabrese S, Gustavsson K, Celani A and Biferale L 2017 {\em Physical Review
  Letters\/} {\bf 118} 158004

\bibitem{Gustavsson2017}
Gustavsson K, Biferale L, Celani A and Colabrese S 2017 {\em The European
  Physical Journal E\/} {\bf 40} 110

\bibitem{Colabrese2018}
Colabrese S, Gustavsson K, Celani A and Biferale L 2018 {\em Physical Review
  Fluids\/} {\bf 3} 084301

\bibitem{Reddy2016}
Reddy G, Celani A, Sejnowski T~J and Vergassola M 2016 {\em Proceedings of the
  National Academy of Sciences\/} {\bf 113} E4877

\bibitem{Reddy2018}
Reddy G, Wong-Ng J, Celani A, Sejnowski T~J and Vergassola M 2018 {\em
  Nature\/} {\bf 562} 236

\bibitem{Biferale2019}
Biferale L, Bonaccorso F, Buzzicotti M, Clark Di~Leoni P and Gustavsson K 2019
  {\em Chaos\/} {\bf 29} 103138

\bibitem{Alageshan2020}
Alageshan J~K, Verma A~K, Bec J and Pandit R 2020 {\em Physical Review E\/}
  {\bf 101} 043110

\bibitem{Monderkamp2022}
Monderkamp P~A, Schwarzendahl F~J, Klatt M~A and L{\"o}wen H 2022 {\em Machine
  Learning: Science and Technology\/} {\bf 3} 045024

\bibitem{Champagne2003}
Champagne L, Carl R~G and Hill R 2003 {\em {Proceedings of the 2003 Winter
  Simulation Conference}\/} {\bf 1-2} 991

\bibitem{Viswanathan1999}
Viswanathan G~M, Buldyrev S~V, Havlin S, da~Luz M~G~E, Raposo E~P and Stanley
  H~E 1999 {\em Nature\/} {\bf 401} 911

\bibitem{Viswanathan2008}
Viswanathan G, Raposo E and da~Luz M 2008 {\em Physics of Life Reviews\/} {\bf
  5} 133

\bibitem{Benichou2005}
B\'enichou O, Coppey M, Moreau M, Suet P~H and Voituriez R 2005 {\em Physical
  Review Letters\/} {\bf 94} 198101

\bibitem{Benichou2006}
B\'enichou O, Loverdo C, Moreau M and Voituriez R 2006 {\em Physical Review
  E\/} {\bf 74} 020102

\bibitem{Loverdo2009}
Loverdo C, B\'enichou O, Moreau M and Voituriez R 2009 {\em Physical Review
  E\/} {\bf 80} 031146

\bibitem{Benhamou1992}
Benhamou S 1992 {\em Journal of Theoretical Biology\/} {\bf 159} 67

\bibitem{Moreau2009}
Moreau M, B{\'{e}}nichou O, Loverdo C and Voituriez R 2009 {\em Journal of
  Statistical Mechanics: Theory and Experiment\/} {\bf 2009} P12006
  
\bibitem{munozgil2023}
Mu{\~n}oz-Gil G, L{\'o}pez-Incera A, Fiderer L~J and Briegel H~J 2023 Optimal
  foraging strategies can be learned and outperform {L}{\'e}vy walks
  (\textit{Preprint} \eprint{2303.06050})

\bibitem{Stanley2002}
Stanley K~O and Miikkulainen R 2002 {\em Evolutionary Computation\/} {\bf 10}
  99

\bibitem{Lang2021}
Lang S, Reggelin T, Schmidt J, M{\"u}ller M and Nahhas A 2021 {\em Expert
  Systems with Applications\/} {\bf 172} 114666

\bibitem{Briegel2012}
Briegel H~J and De las Cuevas G 2012 {\em Scientific Reports} {\bf 2} 1

\bibitem{Song2023}
Song T, Choi Y, Jeon J-H and Cho Y-K 2023 {\em Frontiers in Immunology} {\bf 14} 1129600

\bibitem{Halford2004}
Halford S~E and Marko J~F 2004 {\em Nucleic Acids Research} {\bf 32} 3040

\bibitem{Evans2011}
Evans M~R and Majumdar S~N 2011 {\em Physical Review Letters\/} {\bf 106} 160601

\bibitem{Kusmierz2014}
Kusmierz L, Majumdar S~N, Sabhapandit S and Schehr G 2014 {\em Physical Review Letters\/} {\bf 113} 220602

\bibitem{Kumar2020}
Kumar V, Sadekar O and Basu U 2020 {\em Physical Review E\/} {\bf 102} 052129

\bibitem{Volpe2017}
Volpe G and Volpe G 2017 {\em Proceedings of the National Academy of
  Sciences\/} {\bf 114} 11350

\bibitem{Zanovello2021}
Zanovello L, Caraglio M, Franosch T and Faccioli P 2021 {\em Physical Review
  Letters\/} {\bf 126} 018001

\bibitem{Zanovello2021b}
Zanovello L, Faccioli P, Franosch T and Caraglio M 2021 {\em Journal of
  Chemical Physics\/} {\bf 155} 084901

\bibitem{Zanovello2023}
Zanovello L, L{\"o}ffler R~J~G, Caraglio M, Franosch T, Hanczyc M~M and
  Faccioli P 2023 {\em Scientific Reports\/} {\bf 13} 5616

\bibitem{NEAT}
{\it NEAT-python} \url{https://neat-python.readthedocs.io/en/latest/index.html}

\end{thebibliography}
\end{document}